\documentclass[aps,prd,twocolumn,tightenlines,nofootinbib,superscriptaddress]{revtex4-1}
\usepackage[utf8]{inputenc}
\usepackage{graphicx}
\usepackage{subcaption}
\usepackage{mwe}
\usepackage{mathrsfs}
\usepackage{amsfonts}
\usepackage{setspace}
\usepackage{cellspace}
\usepackage{amsmath,amssymb,bm}
\usepackage[colorlinks=true,linkcolor=blue]{hyperref}
\usepackage{xcolor}
\usepackage{epsfig}
\usepackage{slashed}
\usepackage{caption}
\usepackage{hhline,multirow,tabularx}  
\usepackage{dcolumn}    
\usepackage{url}        
\usepackage{braket}     
\renewcommand{\bra}[1]{\left<#1\left|}
\renewcommand{\ket}[1]{\right|#1\right>}

\begin{document}

\title{\bf{QCD Analysis of Near-Threshold Photon-Proton Production of Heavy Quarkonium}}

\author{Yuxun Guo}
\email{yuxunguo@umd.edu}
\affiliation{Department of Physics, University of Maryland, College Park, MD 20742, USA}

\author{Xiangdong Ji}
\email{xji@umd.edu}
\affiliation{Department of Physics, University of Maryland, College Park, MD 20742, USA}
\affiliation{Center for Nuclear Femtography, SURA, 1201 New York Ave. NW, Washington, DC 20005, USA}

\author{Yizhuang Liu}
\email{yizhuang.liu@alumni.stonybrook.edu}
\affiliation{Institute fur Theoretische Physik, Universitat Regensburg, D-93040 Regensburg, Germany}

\begin{abstract}
The near threshold photo or electroproduction of heavy vector quarkonium off 
the proton is studied in quantum chromodynamics. Similar to the high-energy limit, 
the production amplitude can be factorized in terms of gluonic Generalized 
Parton Distributions and the quarkonium distribution amplitude. At the threshold, the threshold kinematics has a large skewness 
parameter $\xi$, leading to the dominance of the spin-2 contribution over 
higher-spin twist-2 operators. Thus threshold production
data are useful to extract the gluonic gravitational form factors, allowing 
studying the gluonic contributions to the quantum anomalous energy, mass radius, 
spin and mechanical pressure in the proton. We use the recent GlueX data on
the $J/\psi$ photoproduction to illustrate the potential physics impact 
from the high-precision data from future JLab 12 GeV and EIC physics program. 
\end{abstract}
\maketitle

\section{Introduction}

Recently there has been rising interest in measuring the photoproduction of $J/\psi$ particles near the threshold. Dedicated experiments with such purpose are running at Jefferson laboratory (JLab) and similar experiments are planned at the future Electron-Ion Collider (EIC)~\cite{Joosten:2018gyo}. It has been proposed in~\cite{Kharzeev:1995ij,Kharzeev:1998bz} that the photoproduction of $J/\psi$ can be used to measure the gluon matrix element $\langle P|F^2|P\rangle$ in the nucleon and provide crucial information about the trace anomaly contribution to the nucleon mass and mass radius~\cite{Kharzeev:1995ij,Ji:1994av,Ji:1995sv,Ji:2021pys,Kharzeev:2021qkd}. 

In literature, different methods have been adopted to analyze the process. In the vector dominance model~\cite{Kharzeev:1995ij,Kharzeev:1998bz,Kharzeev:2021qkd}, the vector meson photoproduction is related to the forward meson-nucleon scattering where a direct Operator Product Expansion (OPE) in terms of gluonic matrix elements is applicable~\cite{Voloshin:1978hc,Gottfried:1977gp,Appelquist:1978rt,Bhanot:1979vb}. Recently, this process has also been approached using dispersive analysis \cite{Gryniuk:2016mpk,Gryniuk:2020mlh} and holographic Quantum Chromodynamics (QCD)~\cite{Hatta:2018ina,Hatta:2019lxo,Mamo:2019mka,Mamo:2021krl}. However, a comprehensive understanding of the process in perturbative QCD in the threshold region is still lacking except a few early attempts~\cite{Brodsky:2000zc,Frankfurt:2002ka}. This is in sharp contrary to the large $Q^2$ and large $W$ diffractive region where the process has attracts attentions since mid-1990s~\cite{Ryskin:1992ui,Brodsky:1994kf,Frankfurt:1995tf,Frankfurt:1995jw,Frankfurt:1997fj} with well-established all order factorization~\cite{Collins:1996fb} in terms of Generalized Parton Distributions (GPDs)~\cite{Mueller:1998fv,Ji:1996ek,Radyushkin:1996ru}, and the $Q^2=0$, large $W$ and small $t$ region where the factorization has been explicitly shown at next to leading order~\cite{Ivanov:2004vd}. 
In this paper, we study the near threshold photoproduction of heavy vector meson with mass $M_V$ in the heavy-quark mass limit $M_V \rightarrow \infty$. In this limit, the process is dominated by
the direct photon coupling with the heavy quarks, and the heavy quark production through
gluonic subprocesses including possible intrinsic heavy flavor is suppressed by $\alpha_s(M_V)\to 0$. 
We show that the QCD factorization in terms of gluon GPDs in Ref.~\cite{Ivanov:2004vd} remains valid in the threshold region as well. Different from the 
proposals in Refs.~\cite{Kharzeev:1995ij,Kharzeev:1998bz} and recent calculations in a different limit~\cite{Boussarie:2020vmu,Hatta:2021can}, the leading contribution comes from the tensor part of gluonic Energy Momentum Tensor (EMT) and high-dimensional twist-two gluonic operators, due to the emergent light-cone structure in the large $M_V$ limit. The near threshold region is characterized by large skewness parameter $\xi\sim 1$, and the gluon EMT dominates over high dimensional operators
as well as three-gluon exchanges~\cite{Brodsky:2000zc}. Therefore the process can be used to probe the gluonic gravitational form factors of the proton, which provide
important information about the gluon contributions to the proton's mass and spin as well as pressure structures~\cite{Ji:1994av,Ji:1995sv,Polyakov:2002yz,Ji:2020ena,Ji:2021mtz}. 

The organization of the paper is as follows. In Sec.~\ref{sec:kine}, we introduce the near threshold kinematics of the process, paying attention to the emergent light-cone structure in the initial state.  In Sec.~\ref{sec:facto}, we perform the analysis of the two-gluon exchange diagram and express the amplitude in terms of gluon GPDs.  In Sec.~\ref{sec:local}, we Taylor-expand the amplitude in terms of moments of the GPD and show that the cross section can be expressed in terms of the gluonic gravitational form factors of the proton for the large skewness parameter $\xi$ and the decay constant of the heavy meson. In Sec.~\ref{sec:predict}, assuming the dominance of the spin-2
matrix element, we compare our prediction to experiment data of $J/\psi$ production using the results of the gravitational form factors extracted from lattice calculation in~\cite{Shanahan:2018pib} and fit a parametrization of the form factors with the GlueX\cite{Ali:2019lzf} data, especially the $M_A$ and $C(0)$ which are important to the proton mass radius. With those fitted form factors, we predict the cross section of $\Upsilon$ production near threshold. In Sec.~\ref{sec:spin}, we study the polarization effects and make predictions for polarization-dependent cross sections which are important for dis-entangling various form factors. Finally, we make several comments and conclude in Sec.~\ref{sec:conc}. 

\section{Near Threshold Kinematics}\label{sec:kine}

\begin{figure}[t]
    \centering
    \begin{subfigure}[b]{0.45\textwidth}
    \centering
    \includegraphics[width=\textwidth]{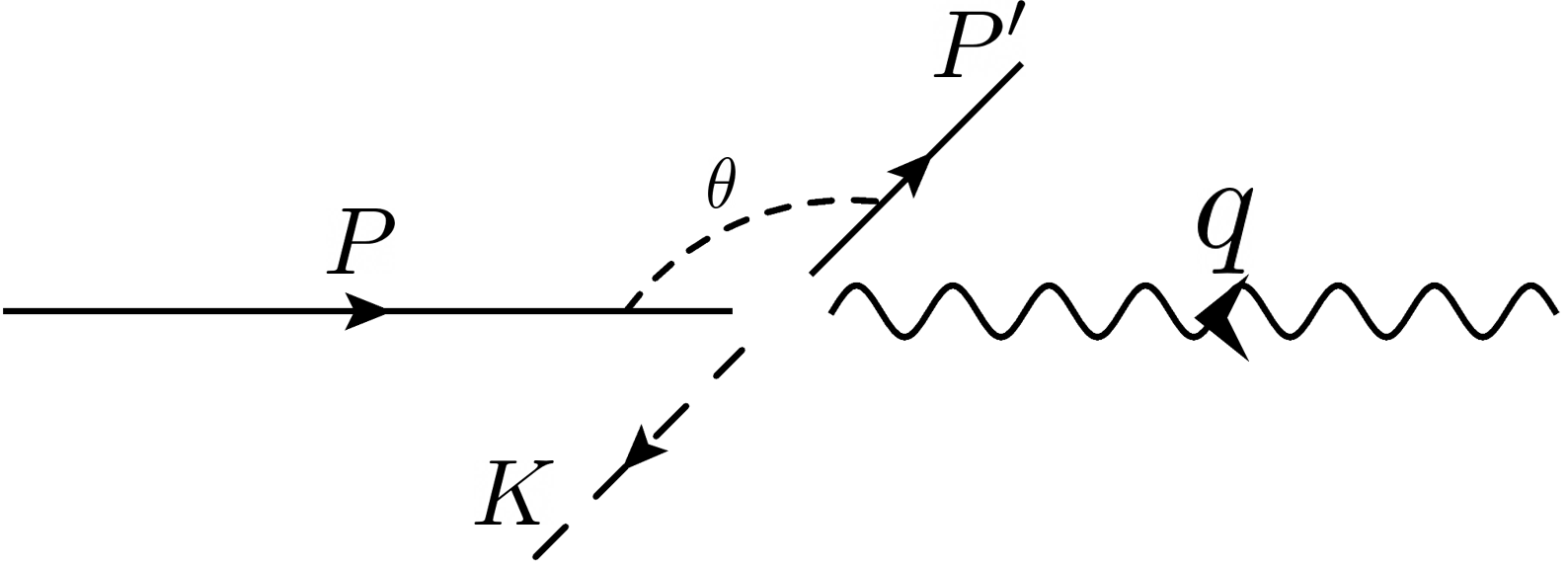}
   \end{subfigure}
 \caption
    {\small The heavy vector-meson photoproduction kinematics in the c.m. frame.}
    \label{fig:Kinamatics}
\end{figure}

We first investigate the near threshold kinematics of the process. Without loss of generality, we work in the center of mass (c.m.) frame though the final result is frame independent. The four-momenta of the incoming photon, outgoing vector meson, incoming proton, outgoing proton are denoted by $q$, $K$, $P$, and $P'$, respectively. We restrict ourselves to the real photon and near threshold region, although similar analysis can be extended to a finite virtual photon mass $Q^2$. We choose the incoming proton to move in the $+z$ direction, as in the lab frame. In c.m. frame, the magnitudes of the three-momenta can be expressed in Lorentz scalars as
\begin{align}
|\vec {K}|&=\sqrt{\frac{\left(W^{2}-\left(M_V+M_N\right)^{2}\right)\left(W^{2}-\left(M_V-M_N\right)^{2}\right)}{4 W^{2}}}\ ,\\
\left|\vec {P}\right|&=\frac{W^2-M_N^2}{2 W}\ ,
\end{align}
where we label the nucleon mass $M_N$, the c.m. energy squared $W^2\equiv (P+q)^2\ge(M_N+M_V)^2$, the average proton four-momentum $\bar P \equiv (P'+P)/2$, four-momentum transfer $\Delta=P'-P$ and
associated invariant $t\equiv \Delta^2$. In the heavy-quark limit, the vector-boson mass $M_V\gg M_N $. As a result $W \gg M_N $ and the incoming proton travels almost along the $+$ light-cone direction in terms of the light-front coordinate $x^{\pm}\equiv \left(x^0\pm x^z\right)/\sqrt{2}$, where $x^\mu$ is
a four-coordinate. One has
\begin{align}\label{eq:lightconeP}
P^\mu \to 2p^\mu \ ,
\end{align} 
with $p=\frac{\bar P^+}{\sqrt{2}}(1,1,0_\perp)$ and $n=\frac{1}{\sqrt{2}\bar P^+}(1,-1,0_\perp)$ being two opposite light-cone unit vectors. The four-momenta of the final-state proton and vector meson are
\begin{align}
K&=(K^0,|\vec K|\vec{e}) \ ,\\
P'&=(P'^0,-|\vec K|\vec{e}) \ ,
\end{align}
where $\vec{e}$ can be in any spatial direction.

\begin{figure}[t]
    \centering
    \includegraphics[width=0.45\textwidth]{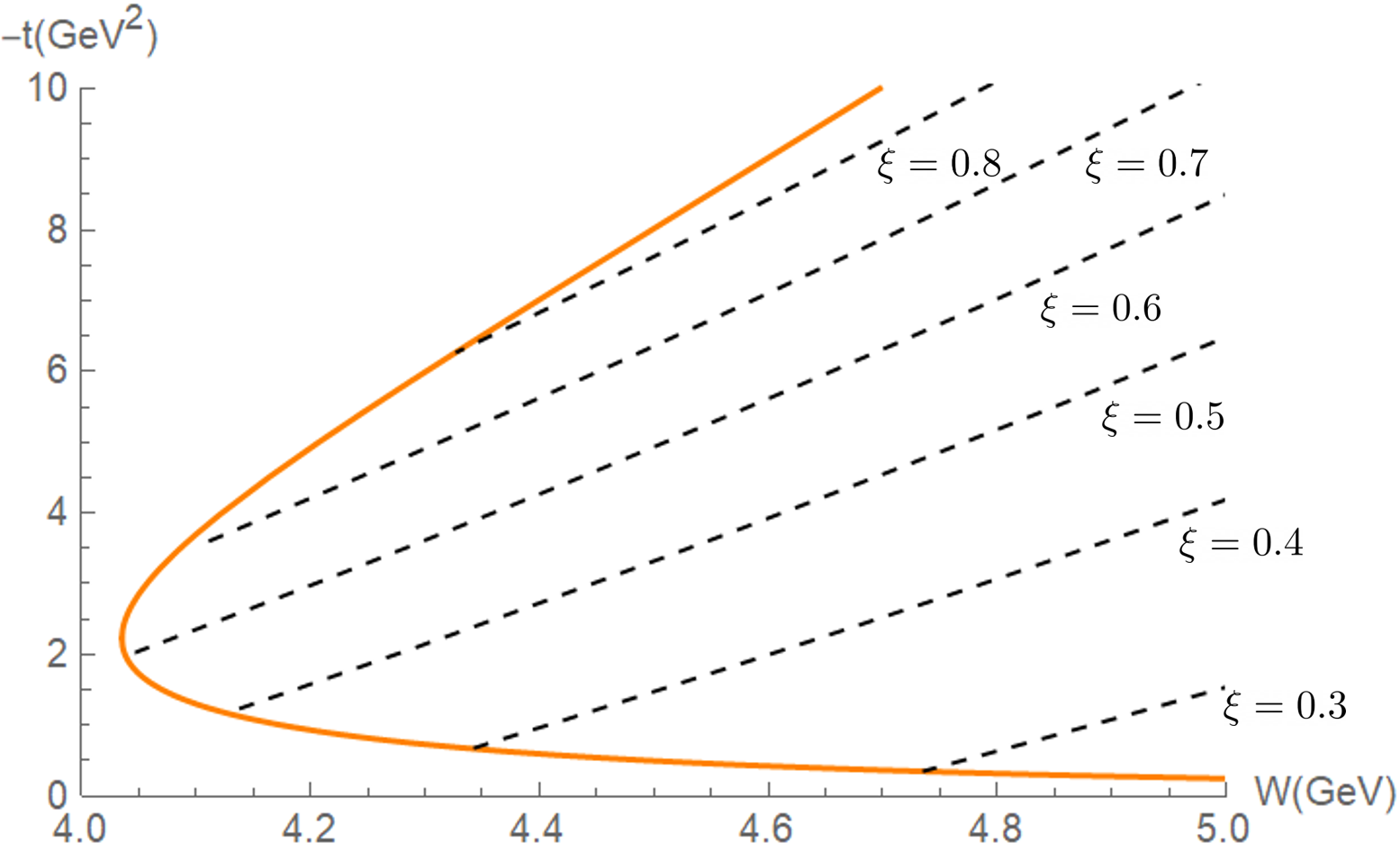}
    \caption
    {\small $\xi$ on the $(W,-t)$ plane in the kinematically allowed region with $M_{J/\psi}=3.097$ GeV.} 
    \label{fig:jpsi}
\end{figure}

\begin{figure}[t]
    \centering
    \includegraphics[width=0.45\textwidth]{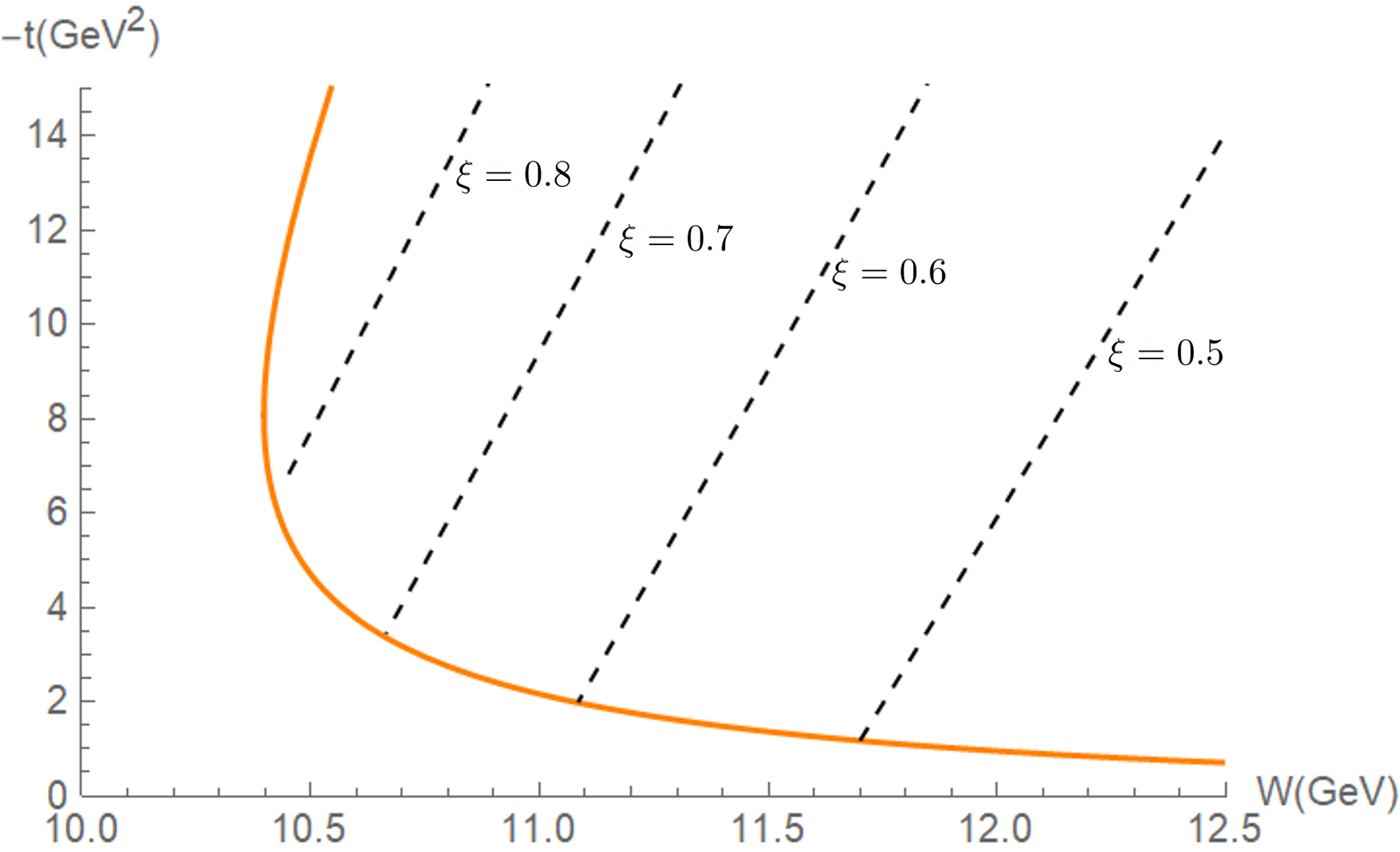}
    \caption
    {\small $\xi$ on the $(W,-t)$ plane in the kinematically allowed region with $M_{\Upsilon}=9.46$ GeV. }
    \label{fig:Upsilon}
\end{figure}

In this paper, we mainly focus on the threshold region defined by the condition that the velocity of the final state proton $\beta\equiv |\vec K|/P'^0$  is of order $1$. 
This condition implies that the velocity of the heavy meson is of order ${\cal O}\left(\frac{M_N}{M_V}\right)$ and $P^+=\frac{M_V}{\sqrt{2}}\left(1+{\cal O}\left(\frac{M_N}{M_V}\right)\right)$. Right at the threshold, the invariant momentum transfer $t$ equals to
\begin{align}
-t_{\rm{th}}=\frac{M_N M^2_V}{M_N+M_V}\ ,
\end{align}
which is of order $M_V M_N$ in the heavy-quark limit. As $W$ increases, the allowed region of $-t$ forms a band $\left[|t|_{\rm min}(W),|t|_{\rm max}(W)\right]$ between the backward case with $-t=|t|_{\rm min}$ and forward case with $-t=|t|_{\rm max}$. Near the threshold, both of them are of order ${\cal O}(M_N M_V)$, much larger than $M_N^2$. In the standard notation of GPDs~\cite{Ji:1996ek}, this implies that the skewness
\begin{align}\label{eq:xith}
    \xi=-\frac{\Delta\cdot n}{2\bar P \cdot n}=\frac{P^+-P'^+}{P^++P'^+}=1+{\cal O}\left(\frac{M_N}{M_V}\right)\ ,
\end{align}
is close to $1$. In fig. \ref{fig:jpsi} and fig. \ref{fig:Upsilon}, we show the $\xi$ value on the $(W,-t)$ plane for $J/\psi$ and $\Upsilon$ production in the kinematically-allowed region. This condition $\xi \to 1$ near threshold will allow
us to study the form factors of the gluon EMT as we discuss later.


\section{Leading-Order QCD Factorization}\label{sec:facto}

\begin{figure}[t]
    \centering
    \begin{subfigure}[b]{0.35\textwidth}
        \centering
        \includegraphics[width=\textwidth]{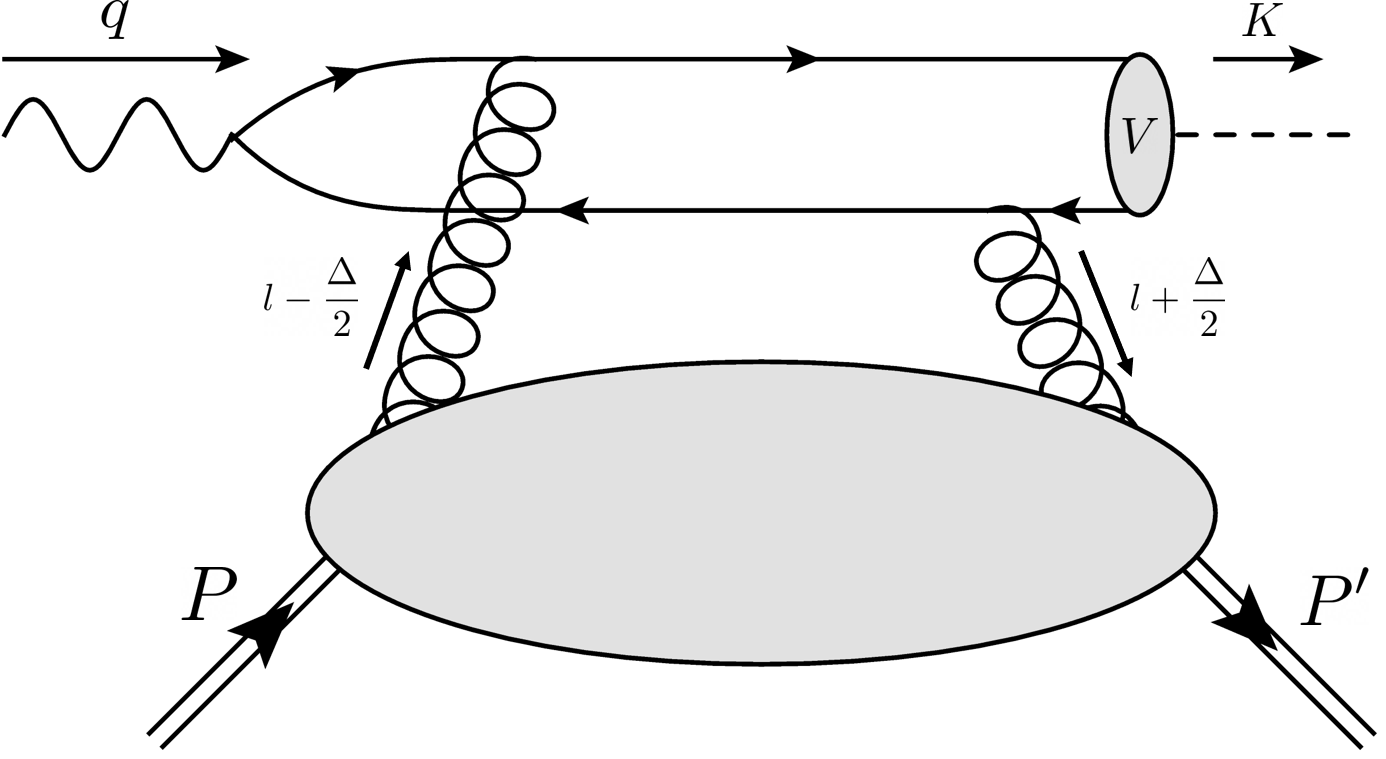}
        \caption{}
        \label{fig:a}
    \end{subfigure}
    \hfill
    \begin{subfigure}[b]{0.35\textwidth}
        \centering
        \includegraphics[width=\textwidth]{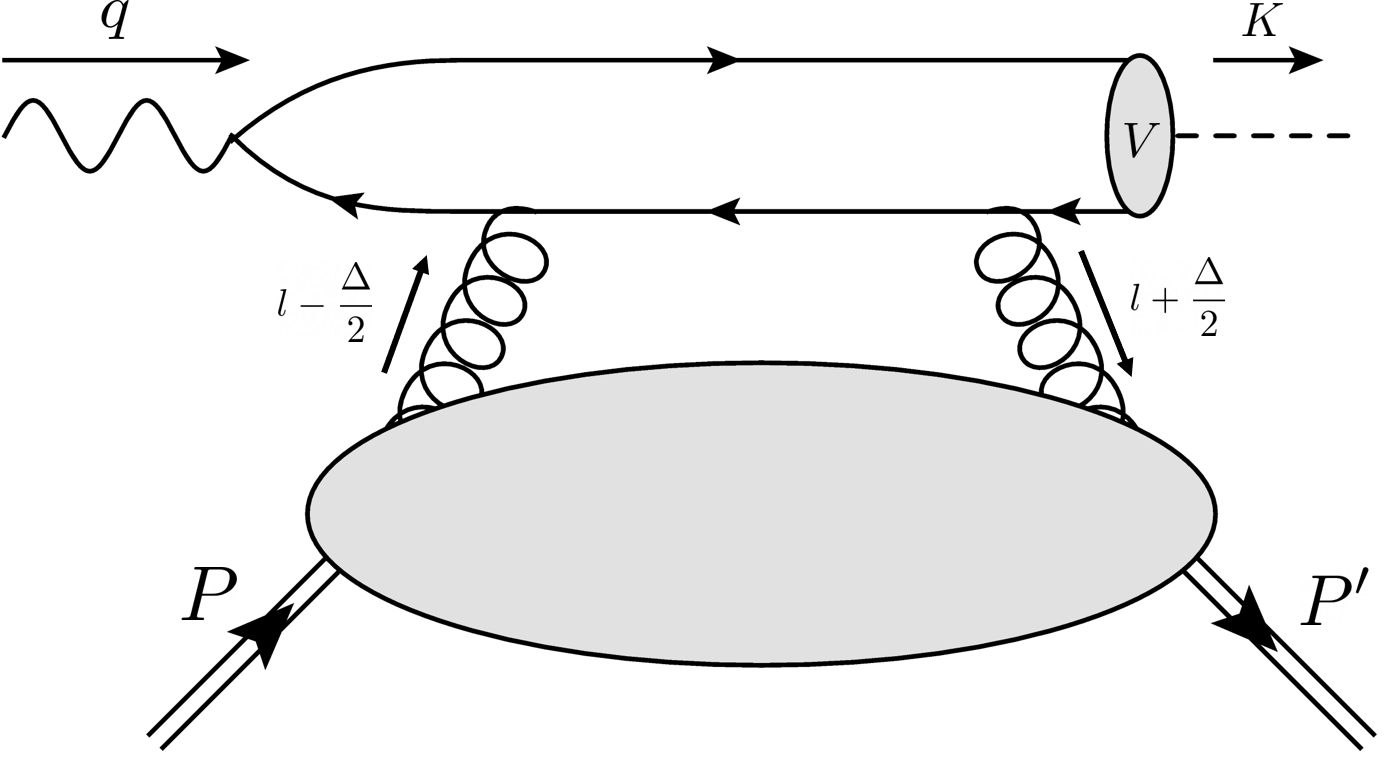}
        \caption{}
        \label{fig:b}
    \end{subfigure}
    \caption
    {\small Examples of leading Feynman diagrams that contribute to heavy vector meson photoproduction.}
    \label{fig:feyndiagram}
\end{figure}

We now proceed to consider the amplitude of the process written as
\begin{equation}
   \mathcal M (q)=\int \text{d}^4 z e^{iq\cdot z} \bra{P',K(\varepsilon_V)} \varepsilon_\mu J^\mu(z) \ket{P} \ ,
\end{equation}
where $J^\mu$ is the electromagnetic current operator, $\varepsilon^\mu$ is the polarization vector of the photon normalized as $\varepsilon^2=-1$ and $\left|K(\varepsilon_V)\right>$ denotes the vector meson state with momentum $K$ and polarization vector $\varepsilon_V$ normalized as $\varepsilon^2_V=-1$ and satisfying $K \cdot \varepsilon_V=0$. All states are normalized
covariantly. In the heavy-quark limit, the leading-order contribution to $\mathcal M$ is given by two-gluon exchange diagrams as illustrated in fig.~\ref{fig:feyndiagram}. To calculate these
contributions, we perform the following approximations which are justified in leading order in ${\cal O}\left(\frac{M_N}{M_V}\right)$ and ${\cal O}\left(\alpha_S\right)$.  

First, since the momentum transfer is mostly in the light-cone direction $p$, only the $+$ components of the loop momenta $l$ and $-\Delta-l$ for the gluons are kept in the heavy-quark loop and can be expressed as
\begin{align}\label{eq:fractions}
-\frac{\Delta}{2}+l&=(\xi+x)p +... \ , \\
-\frac{\Delta}{2}-l&=(\xi-x)p +...\ ,
\end{align}
where $-1<x<1$ is the momentum fraction for $l$.  In the threshold region, $\xi$ approaches $1$ in the heavy-quark limit, though we will keep the $\xi$ dependence in general. 

Second, at the large $M_V$ limit, we perform the non-relativistic approximation to the heavy-meson wave function which amounts to the substitution 
\begin{align}
\langle K(\varepsilon_V)|\psi(-k-K)\bar \psi(k)|0\rangle=\frac{1-\slashed{v}_V}{2}\frac{\slashed{\varepsilon^*_V}}{\sqrt{2}}\frac{1+\slashed{v}_V}{2}\tilde \phi^*(k) \ .
\end{align}
Here $v_V$ is the four-velocity of the vector meson, and the scalar function $\tilde \phi(k)$ corresponds to the distribution amplitude of the vector meson. The relative momentum between the quark and anti-quark is of order ${\cal O}(\alpha_S M_V)$ for hydrogen-like systems, and to the leading order in $\alpha_s$ we can further approximate the wave function as 
\begin{equation}
\label{wfleading}
\begin{split}
    \phi(k)&=(2\pi)^4\delta^4\left(k-\frac{M_V}{2}v_V\right)\int \frac{d^4k'}{(2\pi)^4}\tilde \phi(k')\ , \\ 
&=(2\pi)^4\delta^4\left(k-\frac{M_V}{2}v_V\right)\phi(0)\ ,
\end{split}
\end{equation}
where the velocity $v_V$ of the final state heavy-meson can be approximated as the static one $v=(1,0,0_\perp)$ to leading order in $\frac{M_N}{M_V}$. The $\phi(0)=\sqrt{2M_V/3}\psi_{\rm NR}(0)$ is proportional to the non-relativistic hydrogen-like wave function $\psi_{\rm NR}$ at the origin. The above meson wave function can be studied more systematically with Non-Relativistic QCD (NRQCD), which shows that the leading correction to the wave function is indeed suppressed by order $\alpha_S(M_V)$~\cite{Khan:1995np,Khan:1995pg,Bodwin:2006yd}. 

Finally, we notice that for collinear gluons in light-cone gauge $A^+=0$, the $A^-$ component is suppressed by the large boost factor. Therefore to the leading order approximation, it is sufficient to keep the transverse components of the gluon gauge potential only. 

Given these approximations, we can evaluate the Feynman diagrams in light-cone gauge straightforwardly. The result for the leading amplitude $\mathcal M(\varepsilon_V,\varepsilon)$ reads:
\begin{equation}\label{eq:amplitude1}
\begin{split}
   \mathcal M(\varepsilon_V,\varepsilon)=\frac{8\sqrt{2}\pi \alpha_S(M_V) }{M_V^2 }  \phi^*(0) G(t,\xi)(\varepsilon_V^* \cdot \varepsilon)\ .
\end{split}
\end{equation}
Here the function $G(t,\xi)$ implicitly depends on the polarization of the initial and final proton which is suppressed and can be expressed in terms of the gluon GPDs as 
\begin{align}\label{eq:amplitude2}
G(t,\xi)=\frac{1}{2\xi} \int_{-1}^1\text{d}x{\cal A}(x,\xi)F_g(x,\xi,t) \ .
\end{align}
where the hard kernel ${\cal A}(x,\xi)$ reads
\begin{equation}
   {\cal A}(x,\xi)\equiv\frac{1}{x+\xi-i0}- \frac{1}{x-\xi+i0}\ .
\end{equation}
The standard gluon GPD $F_g$ is defined as~\cite{Ji:1998pc}
\begin{align}
\label{gluonGPD}
   &F_g(x,\xi,t) \equiv\nonumber \\ &\frac{1}{(\bar P^+)^2} \int \frac{\text{d}\lambda}{2\pi} e^{i\lambda x}\bra{P'}\text{Tr}\left\{ F^{+i}_{\;\;\;\;\;}\left(-\frac{\lambda n}{2}\right)F^{+}_{\;\;i}\left(\frac{\lambda  n}{2}\right)\right\} \ket{P}\ ,
\end{align}
where the states are normalized as $\braket{P|P}=2 E_{P} (2\pi)^3\delta^{(3)}(\vec 0)$ and the renormalization scale has been omitted. Similar results as in eq.~(\ref{eq:amplitude1}) and eq.~(\ref{eq:amplitude2}) has been derived in the literature both in the high-energy limit~\cite{Koempel:2011rc,Cui:2018jha} and in the heavy-quark limit~\cite{Ivanov:2004vd}, but not
in the threshold region as we are interested in here.

In eq.~(\ref{eq:amplitude2}), GPDs are evaluated at a generic $\xi$. Near the threshold and in the heavy-quark limit,  $t$ approaches infinity and $\xi$ is close to 1, as shown in eq.~(\ref{eq:xith}). Therefore in principle,
one should set $\xi=1$ our equations above. However, there are two reasons for us to keep the generic $\xi$ dependency. 

First, the expression with generic $\xi$ agrees with the leading order result in the kinematic region at large $W$ and small $|t|$ ~\cite{Ivanov:2004vd,Koempel:2011rc,Cui:2018jha}. What we have shown is that in the heavy-quark limit, the validity region of the leading order factorization formula eq.~(\ref{eq:amplitude1}) and eq.~(\ref{eq:amplitude2}) in terms of GPDs can be smoothly extended to the threshold region along the $|t|_{\rm min}$ line. It shows that the large $M_V$ is sufficient to generate light-cone structure even in the threshold region. Therefore our result can be viewed as a generalization of~\cite{Ivanov:2004vd}. 

Second, although in the heavy meson limit $\xi$ approaches 1 near threshold, in reality, especially the production of $J/\psi$, the kinematic finite meson mass corrections are important. By using the physical $\xi$, one can expect to take part of these kinematic corrections into account. 

Given the amplitude above, we can calculate the cross-section, and the result is
\begin{equation}
\label{xsec}
\begin{split}
\frac{d \sigma}{d t}&=\frac{e^2 e_Q^2}{16 \pi \left(W^2-M_N^2\right)^2} \frac{1}{2}\sum_{\rm polarization}\left| \mathcal M(\varepsilon_V,\varepsilon)\right|^{2}\ ,\\
&= \frac{ \alpha_{\rm EM}e_Q^2}{ 4\left(W^2-M_N^2\right)^2}\frac{ (16\pi\alpha_S)^2}{3M_V^3}|\psi_{\rm NR}(0)|^2 |G(t,\xi)|^2\ ,
\end{split}
\end{equation}
with $e_Q$ the charge of the quark in the unit of proton charge, and the photon and meson polarization are summed over. The kinematic pre-factor is the same as that in~\cite{Hatta:2019lxo,Hatta:2018ina} and in the second line it is shown that the cross section of near threshold meson production is related to the non-relativistic wave function at origin and the gluon GPDs. Again, the renormalization
scales in GPDs and wave function have been omitted
and shall be on the order of $M_V$.

\section{Expansion in Moments of GPD}\label{sec:local}

As shown in eq.~(\ref{eq:amplitude2}), the entire gluon GPD enters in the amplitude for the threshold region heavy quarkonium production because of the light-cone dominance.
However, because $\xi\sim 1$, one can Taylor-expand the ${\cal A}(x,\xi)$ 
and express the $G(t,\xi)$ in terms of the even moments of the GPD     
\begin{align}\label{eq:moments}
    G(t,\xi)=\sum_{n=0}^{\infty}\frac{1}{\xi^{2n+2}} \int_{-1}^{1} dx x^{2n}F_g(x,\xi,t) \ .
\end{align}
The moments can be related to the matrix element of the
gluonic twist-two operator $O_{g}^{\mu\mu_1...\mu_{n-2}\nu}$~\cite{Ji:1998pc}
\begin{align}
    &\langle P'|O_{g}^{\mu\mu_1...\mu_{n-2}\nu}|P\rangle\nonumber \\ 
    &={\cal S} \bar u(P')\gamma^{\mu}u(P)\sum_{i,{\rm even}}A^g_{n,i}(t)\Delta^{\mu_1}..\Delta^{\mu_i}P^{\mu_{i+1}}..P^{\nu} \nonumber \\ 
    &+{\cal S} \bar u(P')\frac{i\sigma^{\mu\alpha}\Delta_{\alpha}}{2m}u(P)\sum_{i,{\rm even}}B^g_{n,i}(t)\Delta^{\mu_1}..\Delta^{\mu_i}P^{\mu_{i+1}}..P^{\nu} \nonumber \\ 
    &+{\cal S}\frac{\Delta^{\mu}}{m}\bar u(P')u(P){\rm mod}(n,2)C_{n}(t)\Delta^{\mu_1}..\Delta^{\nu} \ .
\end{align}
By parametrizing the GPD, 
\begin{align}
    &F_{g}(x,\xi,t)=\frac{1}{2\bar P^+}\times\nonumber \\ &\left[H_g(x,\xi,t)\bar u(P')\gamma^+u(P)+E_g(x,\xi,t)\bar u(P')\frac{i\sigma^{+\alpha}\Delta_{\alpha}}{2m}u(P)\right] \ ,
\end{align}
the moments for the scalar functions $H$ and $G$ can then be written as~\cite{Ji:1998pc}
\begin{align}\label{eq:GPDmoments}
    \int_0^1 dx x^{2n}H_g(x,\xi,t)= \sum_{i=0}^n(2\xi)^{2i}A^g_{2n+2,2i} +(2\xi)^{2n+2}C^g_{2n+2} \ , \\
    \int_0^1 dx x^{2n}E_g(x,\xi,t)= \sum_{i=0}^n(2\xi)^{2i}B^g_{2n+2,2i} -(2\xi)^{2n+2}C^g_{2n+2} \ .
\end{align}
where $C$-terms represent the highest power in $\xi$. In particular, for $n=0$ we have
\begin{align}\label{eq:E0}
    \int_{0}^1 dx H_g(x,\xi,t)=A_{2,0}^g(t)+(2\xi)^2C^g_{2} \equiv H_2(t,\xi)  \ , \nonumber \\
    \int_{0}^1 dx E_g(x,\xi,t)=B^g_{2,0}(t)-(2\xi)^2C^g_{2} \equiv E_2(t,\xi)  \ .
\end{align}
for the leading moments of the GPDs. 

As we will argue, the higher moment contributions
shall be suppressed in the threshold region. 
If one only keep the leading moment $n=0$, the $G(t,\xi)$ becomes
\begin{align}
\label{Gtff}
   G(t,\xi)&=\frac{1}{\xi^2 (\bar P^+)^{2}}\langle P'| \frac{1}{2}\sum_{a,i} F^{a,+i}\left(0\right)F^{a,+}_{\;\;\;\;\;\;i}\left(0\right) |P\rangle  \nonumber \\
   &=\frac{1}{2\xi^2  (\bar P^+)^{2}}\left\langle P^{\prime}\left|T_{g}^{++}\right| P\right\rangle\ ,
\end{align}
in terms of the matrix elements of gluon EMT in the proton.  We then can derive a formula for the cross-section in terms the gravitational form factors defined through following parametrization of the matrix element of gluon EMT~\cite{Ji:1996ek} 
\begin{align}
&\left\langle P^{\prime}\left|T_{q, g}^{\mu \nu}\right| P\right\rangle=\bar{u}\left(P^{\prime}\right)\Bigg{[}A_{q, g}(t) \gamma^{(\mu} \bar{P}^{\nu)}+B_{q, g}(t) \frac{\bar{P}^{(\mu} i \sigma^{\nu) \alpha} \Delta_{\alpha}}{2 M_N}\nonumber \\ 
&+C_{q, g}(t) \frac{\Delta^{\mu} \Delta^{\nu}-g^{\mu \nu} \Delta^{2}}{M_N}+\bar{C}_{q, g}(t) M_N g^{\mu \nu}\Bigg{]} u(P) \ .
\end{align}
where $A=A_{2,0}$, $B=B_{2,0}$ and $C=C_{2,0}$ are the same form factors as in eq.~(\ref{eq:E0}). It leads to the following form of $|G(t,\xi)|^2$ after summing/averaging over the final and initial proton spin
\begin{align}\label{gt2ff}
&|G(t,\xi)|^2=\frac{1}{\xi^4}\Bigg{\{}\left(1-\frac{t}{4M_N^2}\right)E_2^2 \nonumber \\ &-2 E_2(H_2+E_2) +\left(1-\xi^2\right)(H_2+E_2)^2 \Bigg{\}}\ ,
\end{align}
where $H_2\equiv E_2(t,\xi)$, $E_2\equiv E_2(t,\xi)$ are defined in eq.~(\ref{eq:E0}).  Combining with eq. (\ref{xsec}), the cross section of heavy vector meson photoproduction can be expressed in terms of those gravitational form factors. This result agrees with the holographic QCD predictions~\cite{Mamo:2019mka,Mamo:2021krl} that the leading contribution to the cross section is due to exchange of $2^{++}$ excitations, or the spin-2 twist-two operators, instead of the $0^{++}$ $F^2$ operators suggested in~\cite{Kharzeev:1998bz,Kharzeev:2021qkd}. However, in the holographic approach the twist-2 part of the gravitational form factor is dual to the graviton exchange and is free from the $C_g$ contribution. This differs from the generic QCD parametrization in eq.~(\ref{Gtff}). Because of the EMT conservation,
the quantum anomalous energy $F^2$ form factor can be related to the twist-two ones here.   
If the further limit $\xi \rightarrow 0$ is taken in eq.~(\ref{gt2ff}), only the $A$ form factors are leading and all the results agrees. However, this is inconsistent with our approximation here. 

Here we return to the validity of the moment expansion in eq.~(\ref{eq:moments}) and the leading moment approximation in eq.~(\ref{Gtff}). Notice that for generic $\xi<1$, the $G(t)$ has imaginary part \begin{align}
    {\rm Im} \;G(t,\xi) \propto F_g(\xi,\xi,t)+F_g(-\xi,\xi,t)\ ,
\end{align} 
which implies that the moment expansion in eq.~(\ref{eq:moments}) has some 
limitations. However, as $\xi$ gets closer to $1$ in the heavy-quark limit, the above imaginary part vanishes in power of $1-\xi$. The renormalization group evolution will also help to improve the convergence of the expansion at large renormalization scale $\mu\sim M_V$. Indeed, it has been shown that the asymptomatic form of gluon GPD reads~\cite{Goeke:2001tz,Diehl:2003ny}
\begin{align}\label{eq:asym}
F_g^{\rm asym}(x,\xi,t) \propto \left(1-\frac{x^2}{\xi^2}\right)^2\theta\left(1-\frac{x}{\xi}\right) \ ,
\end{align}
which vanishes quadratically at $x=\xi$.  With this form of asymptotic behavior, one can show that the expansion in eq.~(\ref{eq:moments}) is convergent.  To summarize,  eq.~(\ref{eq:moments}) should be a good approximation for $G(t,\xi)$ in the threshold region where $\xi$ is close to $1$.

Regarding the validity of the leading moment approximation in eq.~(\ref{Gtff}), one can reads from eq.~(\ref{eq:moments}) that due to the $\xi^{2n}$ in the denominator, eq.~(\ref{eq:moments}) is simultaneously a moment expansion and a $\xi^{-1}$ expansion. The closer the $\xi$ to $1$, the larger the contribution from the leading term $n=0$.  Using the asymptotic form in eq.~(\ref{eq:asym}), in the exact $\mu \sim M_V \rightarrow \infty$ limit the ratio between the contribution from higher ($n\ge 1$) moments and the leading moment is 
\begin{align}
\frac{\sum_{n\ge 1}^{\infty}\int_{-1}^1 dx \frac{x^{2n}}{\xi^{2n}}F_g^{\rm asym}(x,\xi,t)}{\int_{-1}^1 dx F_g^{\rm asym}(x,\xi,t)}=\frac{1}{4} \ ,
\end{align}
the total higher-moment contribution is 25 percent. 

Realistically, when $\mu \sim M_V$ is not very large and $\xi$ is not exactly at $1$, we can use the 
ratio between the second and the leading terms to estimate the effect of the high-order terms.
Using the explicit expressions for GPD moments in eq.(\ref{eq:GPDmoments}), the ration between second and leading moments can be calculated as 
\begin{align}
    \frac{A^g_{4,0}+4\xi^2A^g_{4,2}+16\xi^4C^g_4}{\xi^2(A^g_{2,0}+4\xi^2C_2^g)}\ ,
\end{align}
for the moments of $H$ and 
\begin{align}
\frac{B^g_{4,0}+4\xi^2B^g_{4,2}-16\xi^4C^g_4}{\xi^2(B^g_{2,0}-4\xi^2C_2^g)}\ ,
\end{align}
for the moments of $E$.  For a quick estimation, one can keep only the $A_0$ form factors which equal to moments of gluon-PDFs at $t=0$. Away from $t=0$, one can use the following dipole model
\begin{align}
    A_{2,0}^g=\left(1-\frac{t}{m^2_A}\right)^{-2} \int_0^1 dx xf_g(x)  \ ,\\ 
    A_{4,0}^g=\left(1-\frac{t}{m^2_{A4}}\right)^{-2} \int_0^1 dx x^3f_g(x) \ .
\end{align}
Here $f_g$ is the gluon PDF at zero momentum transfer and $m_A$, $m_{A4}$ are unknown dipole masses. Neglecting the difference in dipole masses and using the recent CTEQ global analysis~\cite{Hou:2019efy}, at renormalization scale $\mu=1.3$ GeV one has  
\begin{align}
    \frac{\int_0^1 dx x^3f_g(x)}{\xi^2\int_0^1 dx xf_g(x)}=\frac{0.038}{\xi^2} \ .
\end{align}
For $J/\psi$ production, one has $\xi=0.6$ right at the threshold and the ratio is around $0.1$, consistent with the dominance of the leading moment. Away from the threshold, the ratio increases as $\xi$ drops, 
and the dominance of leading moment become less pronounced.  
As one increases the renormalization scale, more momentum fraction will be carried by small $x$ gluons 
and the ratio becomes smaller. However, since the form factor $C$ is also important for the amplitude, a more realistic estimation needs input from lattice calculations.

\section{Gravitational Form Factors from Threshold Data}\label{sec:predict}

As we have discussed in the previous section, threshold quarkonium photoproduction
in the heavy-quark limit is dominated by the nucleon's gravitational form factors in the leading moment approximation. 
In this section, we study the phenomenology of determining these gravitational form factors 
from realistic heavy quarkonium production data, neglecting various higher-order
corrections which we will take into account in the future publication. 
We consider the $J/\psi$ photoproduction where there are more data available near the threshold,  
and predict the near threshold $\Upsilon$ photoproduction cross section 
where the heavy-quark expansion works better.

\subsection{$J/\psi$ Photoproduction}

To start with, we consider the $J/\psi$ photoproduction total cross sections from SLAC \cite{Camerini:1975cy}, Cornell \cite{Gittelman:1975ix} and the most recent GlueX experiments \cite{Ali:2019lzf}. Given that the GlueX data seem to deviate from the SLAC and Cornell data, we will focus on comparing with the GlueX result which includes more data as well as some measurements of differential cross section. 

\begin{figure}[t]
    \centering
    \includegraphics[width=0.45\textwidth]{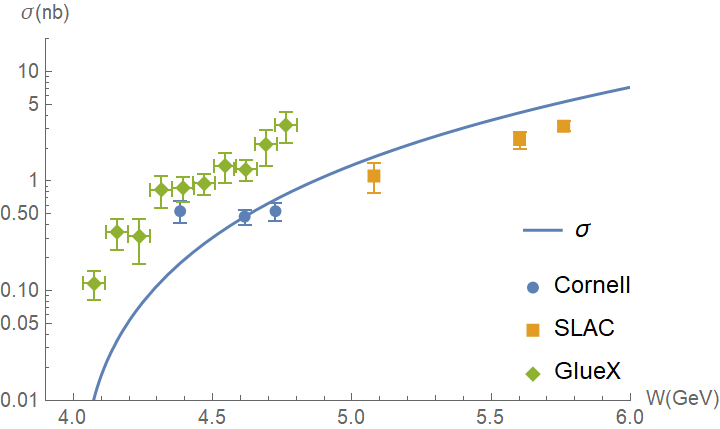}
    \caption
    {\small Comparison of the cross section with gravitational form factors given by lattice\cite{Shanahan:2018pib} with experiments data from SLAC{ \cite{Gittelman:1975ix}}, Cornell{ \cite{Camerini:1975cy}} and GlueX \cite{Ali:2019lzf}.}
    \label{fig:xsec}
\end{figure}

Besides the physical constants such as the proton mass $M_N=0.938$ GeV, the $J/\psi$ mass $M_{J/\psi}=3.097$ GeV etc., we use the same $\alpha_S$ as in \cite{Hatta:2019lxo}
\begin{equation}
    \alpha_S\left(\mu=2\text{ GeV}\right)=0.3\ .
\end{equation}
For the non-relativistic wave function $|\psi_{\rm{NR}}(0)|^2$, it can be measured from the leptonic decay rate of $J/\psi$ \cite{VanRoyen:1967nq,Eichten:1995ch,Bodwin:2006yd}
\begin{equation}
\Gamma\left(V \rightarrow e^{+} e^{-}\right)=\frac{16\pi  \alpha_{\rm EM}^{2} e_{Q}^{2}|\psi_{\rm{NR}}(0)|^2}{M_V^{2}}\left(1-\frac{16 \alpha_{s}}{3 \pi}\right)\ .
\end{equation}
Then one has  \cite{Eichten:1995ch,Eichten:2019hbb}
\begin{equation}
| \psi_{\rm{NR}}(0)|^2= 1.0952 /(4\pi) \left(\text{GeV}\right)^3\ .
\end{equation}
As for $G(t,\xi)$, the input of gravitational form factors is required. At large momentum transfer it has been argued that the form factors decays polynomially~\cite{Brodsky:1973kr,Ji:2003yj,Tong:2021ctu} based on power-counting methods. Here we use the gravitational form factors from lattice calculations, which model those form factors with dipole expansion 
\begin{align}\label{eq:dipole}
A_g(t)&=\frac{A_g(0)}{\left(1-\frac{t}{m_A^2}\right)^2}\ ,\\
C_g(t)&=\frac{C_g(0)}{\left(1-\frac{t}{m_C^2}\right)^2}\ .
\end{align}
In a recent calculation, those parameters are found approximately, $m_A=1.13 \text{ GeV}$, 
$m_C=0.48 \text{ GeV}$, $A_g(0) =0.58$ and $C_g(0)=-1.0$ \cite{Shanahan:2018pib}, while the $B_g(t)$ form factor is numerically small. These inputs without any fitting parameters yield the 
curve shown in fig. \ref{fig:xsec}. 

\begin{figure}[t]
    \centering
    \includegraphics[width=0.45\textwidth]{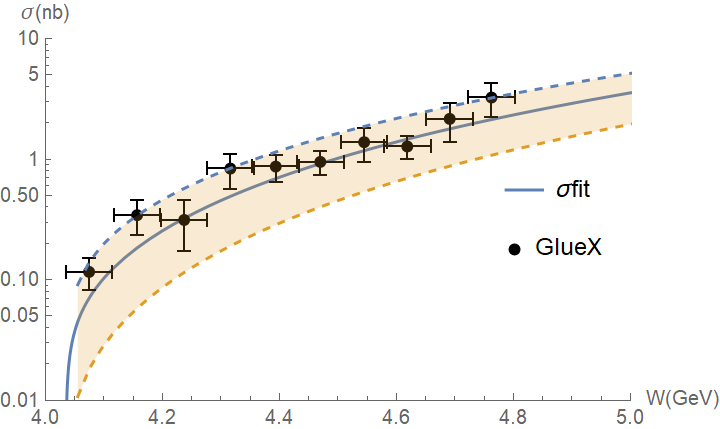}
    \caption
    {\small Fit total cross section compared with the total cross section measured at GlueX \cite{Ali:2019lzf}. The 95\% confidence band is shown as the shaded region hereafter.} 
    \label{fig:fitxsec}
\end{figure}

\begin{figure}[t]
    \centering
    \includegraphics[width=0.45\textwidth]{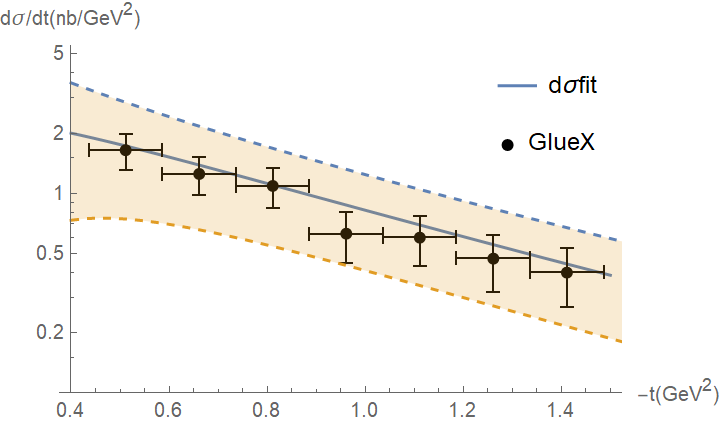}
    \caption
    {\small Fit differential cross section compared with the differential cross section at $W=4.58$ GeV measured at GlueX \cite{Ali:2019lzf}.} 
    \label{fig:fitdifxsec}
\end{figure}

On the other hand, our theoretical formulas 
allow us to extract the gravitational form factors from the $J/\psi$ photoproduction data. 
Since the data are quite limited at this stage, 
We choose to fit the GlueX total cross section and differential cross section combined. 
If we have both $A_g(0)$ and $m_C$ fixed to be the lattice values in order to avoid overfitting, we 
get $m_A=1.64 \pm 0.11 \text{ GeV}$,$C_g(0)=-0.84 \pm 0.82$. In fig. \ref{fig:fitxsec} and fig. \ref{fig:fitdifxsec}, we compare the fit results with the data from GlueX. The uncertainties of $m_A$ and $C_g(0)$ indicate that the data are more sensitive to $m_A$ rather than $C_g(0)$ in the region of GlueX data. One can of course explore other fitting scenarios as well. 


\subsection{$\Upsilon$ Photoproduction}

\begin{figure}[t]
    \centering
    \includegraphics[width=0.45\textwidth]{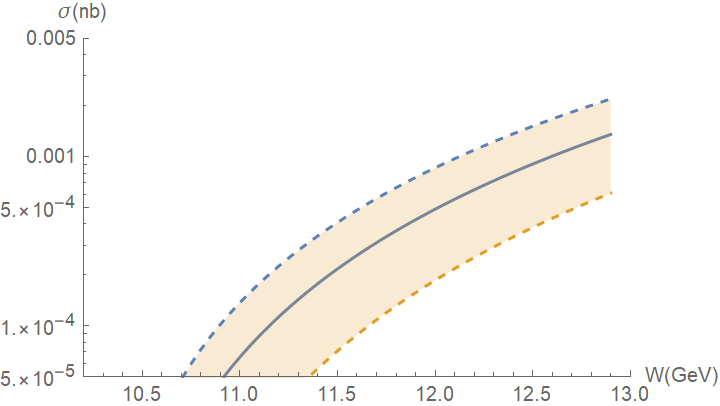}
    \caption
    {\small Solid line shows the predicted total cross section for $\Upsilon$ photoproduction near threshold.} 
    \label{fig:xsecupsilon}
\end{figure}

\begin{figure}[t]
    \centering
    \includegraphics[width=0.45\textwidth]{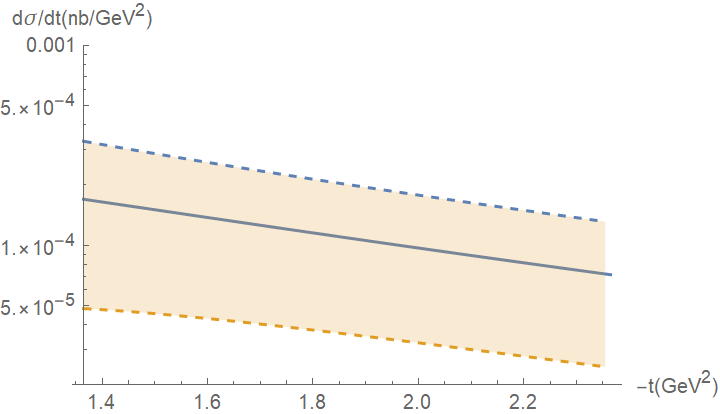}
    \caption
    {\small Solid line shows the predicted differential cross section at $W=11.5$ GeV for $\Upsilon$ photoproduction near threshold.} 
    \label{fig:difxsecupsilon}
\end{figure}

The above result can be used to predict the $\Upsilon$ photoproduction rate near threshold with $M_{\Upsilon}=9.46$ GeV and the wave function at origin for $\Upsilon$
 \cite{Eichten:1995ch,Eichten:2019hbb}
\begin{equation}
| \psi_{\rm{NR}}(0)|^2= 5.8588 /(4\pi) \left(\text{GeV}\right)^3\ .
\end{equation}
Considering the effect of running coupling constant, $\alpha_S = 0.2$ is used for the $\Upsilon$ production. Then we have the predicted total cross section as shown in fig. \ref{fig:xsecupsilon} and the predicted differential cross section in fig. \ref{fig:difxsecupsilon}. The large mass of $\Upsilon$ implies that our calculation in the heavy meson limit works better for $\Upsilon$ production, as there are less theoretical uncertainties from higher order correction. However, the production rate is suppressed by the heavy meson mass and thus the cross section is much lower than $J/\psi$.

\subsection{Quantum Anomalous Energy, Mass Radius, and Pressure distribution}

The gravitational form factors measured from $J/\psi$ photoproduction can be used to study mass, spin and 
pressure properties of the nucleon~\cite{Ji:1994av,Ji:1996ek,Polyakov:2002yz,Kharzeev:2021qkd,Ji:2021mtz}. 

There are two methods to access the quantum anomalous energy through the matrix elements of the EMT. The first is through the matrix element of $F^2$ suggested in~\cite{Kharzeev:1995ij}. In our factorization, this is a sub-leading contribution. Alternatively, 
one can also access the same quantity
by considering the trace of the full EMT, 
\begin{eqnarray}
      H_a &=& \langle P|T^\mu_{~\mu}|P\rangle \\
          &=& \frac{1}{4} (A_q(0)+A_g(0))M_N \ , 
\end{eqnarray}
where $A_q(0)$ and $A_g(0)$ are
traditionally interpreted as the momentum fractions
carried by quarks and gluons, respectively.
The above relation comes from the conservation
of the EMT. Thus fitting $A_g(0)$ through  heavy-quarkonium production data, one obtains 
the gluonic contribution to the quantum
anomalous energy, and the result can be compared with the vector-dominance model analysis~\cite{Kharzeev:1995ij,Wang:2019mza}. Of course, it also provides an alternative 
determination of the gluonic momentum fraction
in the nucleon. It shall be clear that our formula 
for the cross section works for the large $t$. 
To extrapolate to $t=0$, one has to learn the 
$t$ dependence through lattice QCD calculations.

The scalar and mass radii of the proton are
defined as~\cite{Ji:2021mtz}, 
\begin{equation}
\begin{aligned}
\left\langle r^{2}\right\rangle_{s} &=6 \frac{d A\left(t\right)}{d t}-18 \frac{C(0)}{M_N^2}\ , \\
\left\langle r^{2}\right\rangle_{m} &=6 \frac{d A\left(t\right)}{d t}-6 \frac{C(0)}{M_N^{2}}\ .
\end{aligned}
\end{equation}
The two different mass radii are related with the $C(0)$ term as
\begin{equation}
\left\langle r^{2}\right\rangle_{s}-\left\langle r^{2}\right\rangle_{m}=-12\frac{C(0)}{M_N^{2}}\ .
\end{equation}
The gravitational form factors here are for the total EMT so one has $A(t)=A_{q}(t)+A_g(t)$ and $C(t)=C_{q}(t)+C_g(t)$. For the quark contribution we use the lattice data $C_{u+d}(0)=-0.267$ or $C_{u+d}(0)=-0.421$ depending on the extrapolation method~\cite{Hagler:2007xi}. As for $A_{u+d}(t)$, one could also fit the form factor with above dipole expansion as 
\begin{align}
A_{u+d}(t)&=\frac{A_{u+d}(0)}{\left(1-\frac{t}{m_{{\rm Di},u+d}^2}\right)^2}\ ,
\end{align}
for which one has the dipole mass $m_{{\rm Di},u+d} \approx 1.8$ GeV and $A_{u+d}(0)\approx 0.5$~\cite{Hagler:2009ni}. Then the quark form factors combined with above gluon form factors from fitting gives
\begin{align}
\left\langle r^{2}\right\rangle_{A}&\equiv  6 \frac{d A\left(t\right)}{d t}\approx (0.42~ \rm{fm})^2\ , \\
\left\langle r^{2}\right\rangle_{C}&\equiv -6 \frac{C(0)}{M_N^2}\approx(0.54~\rm{fm})^2\ .
\end{align}
and then we have 
\begin{align}
\left\langle r^{2}\right\rangle_{s} &\approx (1.03~\rm{fm})^2\ ,\\
\left\langle r^{2}\right\rangle_{m} &\approx (0.68~\rm{fm})^2\ .
\end{align}
One should be aware that those results are associated with large uncertainties resulting from the lack of precision for the fitted value $C_g(0)=-0.84 \pm 0.82$. 

\begin{figure}[t]
    \centering
    \includegraphics[width=0.45\textwidth]{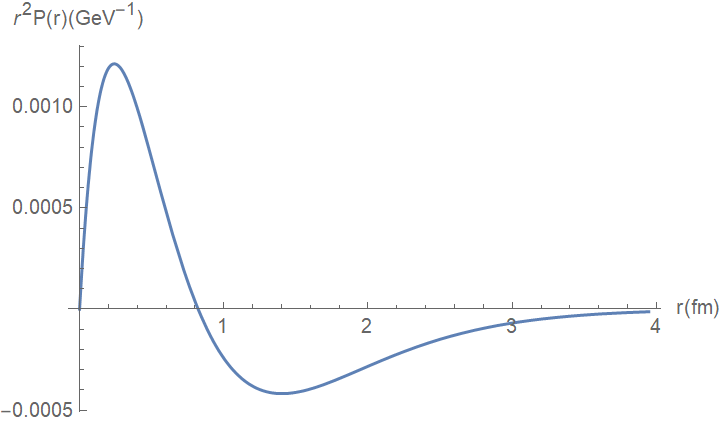}
    \caption
    {\small The gluon contribution to the pressure distribution inside the proton from the combined lattice calculation~\cite{Shanahan:2018nnv} and out fit to GlueX data.} 
    \label{fig:pressure}
\end{figure}

The gravitational form factor $C$ also provides a direct access to the pressure and shear force distributions of the nucleon~\cite{Polyakov:2002yz,Polyakov:2018zvc,Burkert:2018bqq,Shanahan:2018nnv}. 
Given our fit value $C_g(0)=-0.84 \pm 0.82$ and combining with the lattice data $m_C=0.48$ ${\rm GeV}$ in the dipole assumption eq.~(\ref{eq:dipole}), the gluon contribution to the pressure distribution is shown as fig. \ref{fig:pressure}. Here we have neglected the $\bar C$ contribution. 

Above results serve as an example of measuring the gravitational form factors from the photoproduction of heavy vector mesons. Although the results rely on the input from lattice calculations as well as the models for the form factors, it can be a useful tool when more data are available and higher-order corrections are taken into account. One can also
make a full GPD-based analysis directly using Eq. (14). 

\section{Cross section with Polarization}\label{sec:spin}

Up to now we have only considered the unpolarized cross section. In this section we study the polarization effect. By measuring the polarization of the photon and the vector meson, we are able to verify the specific structure $\varepsilon_V^* \cdot \varepsilon$ of the factorization formula eq.~(\ref{eq:amplitude1}) which will serve as a crucial test of the factorization formula itself. Since eq.~(\ref{eq:amplitude1}) also predicts that the dependence on the proton polarization is included in the matrix element $G(t,\xi)$, therefore by measuring the polarization of the proton also allow us to obtain more information to extract the gravitational form factors. Notice that the measurement of the polarization of the final state proton can be non-trivial~\cite{Gayou:2001qd,Gayou:2001qt}. In this section we shall study the polarized cross section from two different point of view, the vector meson polarization and the proton polarization.

Polarization observables allow us to separate out the contributions
from different form factors. In particular, even though the $B_{q,g}(t)$ 
is small, the polarized cross sections can be sensitive to the $B_g(t)$ form factor which is related to the gluon angular momentum~\cite{Ji:1996ek}. 
If given sufficient data with high precision, it might be possible to 
shed some light on the gluon angular momentum contribution to the proton spin.

\subsection{Vector Meson Polarization}
The total cross section is proportional to $|\mathcal M(\varepsilon_V,\varepsilon)|^2$ and thus one has from eq. (\ref{eq:amplitude1}),
\begin{equation}
    \frac{d\sigma}{dt} \propto \varepsilon^*_\mu(q,\lambda) \varepsilon_\nu(q,\lambda) \varepsilon^{*\mu}_V(K,\lambda') \varepsilon^\nu_V (K,\lambda')\ .
\end{equation}
In general one could define the photon polarization tensor and meson polarization as
$\rho_{\mu\nu}(q)$ and $\rho_{V\mu\nu}(K)$ so $ \frac{d\sigma}{dt} \propto\rho_{\mu\nu}(q)\rho_{V}^{\mu\nu}(K)$. For physical photon or meson, the polarization tensor is simply 
\begin{align}
   \rho_{\mu\nu}(q) &= \varepsilon^*_\mu(q,\lambda) \varepsilon_\nu(q,\lambda)\ ,\\
    \rho_{V\mu\nu}(K) &= \varepsilon^*_{V,\mu}(K,\lambda') \varepsilon_{V,\nu}(K,\lambda')\ ,
\end{align}
while in the case they are virtual particle coupled with leptonic current, the polarization tensor can be written with those  leptonic current as
\begin{align}
   \rho_{\mu\nu}(q) = \bar u(l_1,S_1) \gamma_{\mu} u(l_2,S_2) \bar u(l_2,S_2) \gamma_{\nu} u(l_1,S_1)\ ,
\end{align}
with $l_i$ and $S_i$ the momenta and the polarization vectors of the lepton and anti-lepton pair. They satisfies $S_i \cdot l_i=0$, $S_i^2=-1$ and similarly for the vector meson. If we consider the photoproduction process when the photon is on-shell and unpolarized, the polarization tensor can be written as
\begin{align}
   \rho_{\mu\nu}(q) = \frac{1}{2}g_{\mu i}g_{\nu j}\left(\delta^{ij}-\hat q^i\hat q^j\right)\ ,
\end{align}
where $\hat q^i$ is the unit vector in the direction of photon three-momentum. As for the vector meson, assuming the final leptons are unpolarized and averaging over the lepton polarization we have 
\begin{align}
   \rho_{V\mu\nu}(K) = -\frac{1}{2}K^2 g_{\mu\nu}+l_{1,\mu}l_{2,\nu}+l_{2,\mu}l_{1,\nu}\ .
\end{align}
with $l_1,l_2$ the momenta of the two leptons satisfying $l_2+l_1=K$, and thus we have 
\begin{equation}
    \frac{d\sigma}{dt} \propto \frac{K^2}{2} +\vec l_1 \cdot \vec l_2- \hat q\cdot \vec l_1 \hat q\cdot \vec l_2\ .
\end{equation}
The last term $\hat q\cdot \vec l_1 \hat q\cdot \vec l_2$ indicates that the cross section depends on the angle between the momentum of outgoing lepton and the momentum of initial photon and can be examined by experiments.

\subsection{Proton Polarization}

Here we consider the polarization of the initial and final proton. For a demonstration, let us consider the asymmetric-difference $\delta |G(t,\xi,S',S)|^2$ as
\begin{equation}
    \begin{split}
       \delta | G(t,\xi,S',S)|^2 \equiv|G(t,\xi,S',S)|^2 -|G(t,\xi,S',-S)|^2\ ,
    \end{split}
\end{equation}
which measures the difference in the cross section when the spin $S^\mu$ of the initial proton has been flipped. Notice the polarization vector satisfies $S \cdot P =0$ and $S^2=-1$ and similarly for $S'$. Then one has 
\begin{widetext}
\begin{equation}
\label{deltag}
    \begin{split}
      \delta | G(t,\xi,S',S)|^2=\frac{1}{\xi^4}\Bigg{\{}&-S\cdot S'\bigg{[}
      \left(1-\frac{t}{4M_N^2}\right)E_2^2-2(E_2+H_2) E_2 +(1-\xi^2)(E_2+H_2)^2\bigg{]}\\
       &+\frac{S'^+ (P'\cdot S)}{\bar P^+} \left[H_2+E_2\right]\left[\mathcal (1+\xi)H_2+\xi E_2\right]+ \frac{(P\cdot S')(P'\cdot S)}{2M_N^2}E_2^2 \\
       &+\frac{S^+(P\cdot S')}{\bar P^+}\left[H_2+E_2\right]\left[ (1-\xi)H_2-\xi E_2\right]+ \frac{S^+ S'^+ t}{2(\bar P^+)^2}(H_2+E_2)^2\Bigg{\}} \ ,
    \end{split}
\end{equation}
where $E_2\equiv E_2(t,\xi)$ and $H_2\equiv H_2(t,\xi)$ are leading moments of GPDs defined in eq.~(\ref{eq:E0}).
When the initial proton are transversely polarized, one can show that
\begin{equation}\label{eq:crossTr}
    \begin{split}
    &|G(t,\xi,S',S)|^2=\frac{1-S\cdot S'}{4}\sum_{S,S'}|G(t,\xi,S,S')|^2 + \frac{S'^+ (P'\cdot S)}{2\xi^4\bar P^+}\left[H_2+E_2\right]\left[\mathcal (1+\xi)H_2+\xi E_2\right]+\frac{(P\cdot S')(P'\cdot S)}{4\xi^4 M_N^2}E_2^2 \ .
\end{split}
\end{equation}
\end{widetext}
In particular, if both $S$ and $S'$ are perpendicular the scattering plane, then only the first term in eq.~(\ref{eq:crossTr}) survives and the cross section is non-zero only when $S$, $S'$ are parallel. For generic transverse spin $S$, $S'$ the second term is also non-zero. By measuring the cross section at different $S$, $S'$, one is allowed to extract all the three form factors. 

\section{Comments and Conclusion}\label{sec:conc}

In this paper, we have made a heavy-quark 
expansion for the vector meson production at
the threshold region in QCD. The result
shows that the cross section can be
used to measure the form factors of 
the gluonic EMT. Before ending the paper, we would like to make several comments: 

First, our calculation is performed only at leading order in ${\cal O}\left(\frac{M_N}{M_V}\right)$ and ${\cal O}(\alpha_S)$. For application to $J/\psi$ production, since the $J/\psi$ mass is not heavy enough, one expects large mass correction and high-twist effect. A comprehensive study of the mass correction is definitely very important but is beyond the scope of the current paper. Beyond leading order in ${\cal O}\left(\alpha_S\right)$, there are quantum corrections both from the emission of virtual gluon and from the internal motion of the vector meson~\cite{Ivanov:2004vd}. For production of $J/\psi$, the order ${\cal O}(\alpha_s)$ corrections can be significant~\cite{Ivanov:2004vd}. 
 A calculation of these effects will be left to a future work.

Second, we should emphasize again that our result should be viewed as a generalization of the leading order factorization in~\cite{Ivanov:2004vd}. The kinematic region considered therein is large $W$ and small momentum transfer $t$, corresponding to the large $W$ part of the $t_{\rm min}$ line. The heavy-quark limit is performed after taking the large $W$ limit. In this work we have extended the factorization along the $t_{\rm min}$ line into the entire threshold region where $W$ is of the same order of $M_V$ and is a single ultra-violet scale. While the factorization seems to work in leading and next leading order, the validity of the factorization theorem to all orders in perturbation theory remains to be established. 

Third, although in this paper we considered only the small $Q^2\sim 0$ region, at large $Q^2$ the factorization formula in terms of GPD should remain valid. In fact, as $Q^2$ increase, the incoming proton becomes faster and the momentum transfer becomes larger. The light-cone structure shall be more relevant. This picture differs qualitatively from the Euclidean OPE approach in Ref.~\cite{Boussarie:2020vmu}.

Forth, we comment on the scalar contributions. In eq.~(\ref{eq:amplitude1}), only the $F^{+i}F^{+i}$ component has been kept due to the standard power-counting rule of collinear gluon. If one keep all the terms, then the $F^{+-}F^{+-}$ contribution, corresponding to the $F^2$ contribution is proportional to $1-\xi^2$ and is suppressed in the heavy-quark limit.  A detailed study of $F^2$ and three-gluon contribution is left to future work.

Finally, one should also notice that although in the current work only two-gluon exchange has been considered, starting from twist-$3$ there can be three-gluon contributions as well. It has been argued in~\cite{Brodsky:2000zc} that in the near threshold region the three gluon contributions might be important. To further clarify the importance of the three and more gluon contributions requires a throughout power-counting analysis in the threshold region. A throughout study of three-gluon contribution will be left to a future work. 
 
In conclusion, we have shown that the factorization formula in terms of gluon GPD for the photo or elector production of heavy vector meson in ~\cite{Ivanov:2004vd} remains to be valid in the near threshold region. At a large vector meson mass, the skewness $\xi$ is close to $1$ and the amplitude is dominated by the leading moment of the gluon GPD, the gluon EMT or gravitational form factors of the proton. This allows us to extract alternatively the gluonic contribution to the momentum and anomalous quantum energy. It also allows to extract 
the gluonic $C$ form factor (or $D$ terms) and determine the mass radius and pressure 
distribution of the proton.  We have applied our formalism to the case of unpolarized $J/\psi$ production with GlueX data, and the result is encouraging. 

Acknowledgment: We thank Y. Hatta, D. Kharzeev, J. P. Ma, A. Sch\"afer, F. Yuan, and I. Zahed for useful discussions and correspondences. We particularly thank Z.-E.Meziani for inspiring us to study heavy-quarkonium 
threshold production and polarization observables. This research is partly supported by the U.S. Department of Energy, Office of Nuclear Physics, under contract number DE-SC0020682, and by Southeastern Universities Research Association (SURA). Y. Guo is partially supported by a graduate fellowship from Center for Nuclear Femtography, SURA, Washington DC.

\bibliographystyle{apsrev4-1}
\bibliography{refs.bib}
\end{document}